\documentstyle[preprint,prc,aps,floats,epsf]{revtex}

\tightenlines

\newcommand{\beq}{\begin{equation}}
\newcommand{\eeq}{\end{equation}}
\def\be{\begin{eqnarray}}
\def\ee{\end{eqnarray}}
\newcommand{\ben}{\begin{eqnarray*}}
\newcommand{\een}{\end{eqnarray*}}

\def\simlt{\stackrel{<}{{}_\sim}}

\newcommand{\wt}{\widetilde}
\def\({\left( }
\def\){\right) }
\def\[{\left[ }
\def\]{\right] }

\newcommand{\order}{{\cal O}}

\newcommand{\dsl}{\raise.15ex\hbox{$/$}\kern-.57em\hbox{$\partial$}}
\newcommand{\Dsl}{\raise.15ex\hbox{$/$}\kern-.65em\hbox{$D$}}

\newcommand{\kf}{k_F}
\newcommand{\as}{a_s}
\newcommand{\kappab}{\mbox{\boldmath$\kappa$}}
\newcommand{\cD}{{\cal D}}

\begin{document}
\draft

\preprint{MIT-CTP-3033}
\title{Effective Field Theory Power Counting at Finite Density}

\author{James V. Steele}
\address{Center for Theoretical Physics and 
Laboratory for Nuclear Science\\
Massachusetts Institute of Technology, Cambridge, MA\ \ 02139, USA}

\date{November 9, 2000}

\maketitle

\begin{abstract}
Effective field theory is applied to 
finite-density systems 
with two-particle interactions exhibiting
an unnaturally large scattering length, such as neutron matter. 
A new organizational scheme, identified 
for a large number of space-time dimensions,
allows for convergent analytic calculations 
in many-body systems and is similar to 
the hole-line expansion of traditional nuclear physics. 
\end{abstract}

\thispagestyle{empty}
\pacs{PACS number(s): 21.65.+f, 24.10.Cn, 11.10.-z}

\newpage



Effective field theory (EFT) 
provides a way to systematically improve low-energy observables
and make predictions for related processes.
It has recently been applied to 
systems with an unnaturally large scattering length
by adopting an organizational scheme that 
forces an infinite set of diagrams to be summed
at leading order~\cite{Weinberg:1991um}.

Extending 
the systematic language of an EFT to many-body systems
would be beneficial for describing 
various systems of interest including 
nuclei, neutron stars, and
atomic Fermi-Dirac condensates. 
Although sophisticated approximation techniques 
already exist 
for describing finite-density systems~\cite{Negele,wiringa},
error estimates are difficult to quantify and have been 
considered 
the ``holy grail'' of many-body physics by Brueckner. 

At finite density, EFTs have given an accurate understanding of 
dilute systems~\cite{Hammer:2000xg} and of
Fermi surface
effects~\cite{Polchinski:1992ed}
such as the superconductor pairing gap~\cite{Papenbrock:1999wb,Evans:1999ek}.
Also, the Skyrme model and EFTs constructed at saturation
density~\cite{Lutz:2000vc} can parameterize finite-density effects with
only a few parameters.  
However, connecting bulk properties
at finite density, 
such as the equation of state of nuclear matter
(an idealized infinite system of protons and neutrons interacting only
through the strong force), 
to known free-space interactions between the constituent
particles requires a consistent organizational scheme (or power
counting) at finite density. 
This will also allow many other interesting
questions to be investigated, including 
whether nuclear matter binds in the chiral
limit~\cite{Bulgac:1997ji} and why nuclear matter saturates.

An EFT treatment of many-body systems automatically dictates the relevance 
of higher-particle contact interactions. 
For example, three-particle scattering composed of only two-body
interactions has divergent loop corrections and the addition of
an actual three-body
contact interaction is required to regularize the 
result~\cite{Braaten:1996rq}.
For nucleons in the triton, such a 
contact interaction has been shown to be as important as two-body
interactions~\cite{Bedaque:2000ve}. 
Therefore, it is natural to first focus on an EFT description of 
finite density systems with two states, such as
neutron matter, since then
Pauli's exclusion principle requires three- and
higher-body contact interactions to 
be multiplied by powers of momentum
and the EFT 
implies these contributions are suppressed.

For neutrons, restricting the momenta to be below $m_\pi$
and neglecting spin dependence
will clarify the discussion since pions and other extended 
structure can be integrated out leaving only simple point-like
interactions.
In addition, the two proposed free-space EFT power countings,
known as $\Lambda$-counting and 
$Q$-counting~\cite{Weinberg:1991um}, yield equivalent results 
in this regime.
Therefore, $\order(\kf/m_\pi)$ corrections
will be implicit in the following.
To make the relevant scales transparent,
dimensional regularization with power divergence
subtraction (PDS)~\cite{Kaplan:1998we} will be used below to regulate
infinite integrals. 

\begin{figure}[t]
\begin{center}
\leavevmode
\epsfysize=1in
\epsffile[255 541 376 628]{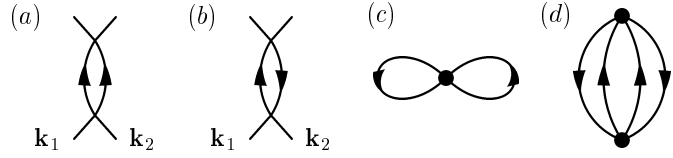}
\end{center}
\caption{\label{fig1} Scattering with propagators in the same
direction (a) and opposite directions (b).  Also shown are the two
simplest contributions to the ground-state energy (c) and (d).
} 
\end{figure}

Non-relativistic particles with (spin-averaged) contact interactions
are governed by the most general lagrangian 
\be
{\cal L}_{\rm eff} &=& N^\dagger \left( i \partial_t + \frac{\nabla^2}{2M}
\right) N
- C_0 (N^\dagger N)^2 + \frac12 C_2 \left[ (N^\dagger
\nabla N)^2 + ((\nabla N^\dagger) N)^2 \right]
\nonumber\\
&&{}+ C_2^p ( N^\dagger \nabla N) \cdot \( (\nabla N^\dagger) N\)
 + \ldots \, ,
\label{Leff}
\ee
with particle mass $M$, 
$S$-wave interactions $C_0$ and $C_2$, $P$-wave interaction
$C_2^p$, and higher derivatives and partial waves represented 
by the dots.
Calculating observables using this non-renormalizable lagrangian 
leads to divergences which need to be regulated.
One-loop scattering 
(Fig.~\ref{fig1}a) with external
relative momentum ${\bf k}=\frac12 ( {\bf k}_1 - {\bf k}_2)$ in the
PDS subtraction scheme is 
\beq
\int\!\! \frac{d^3q}{(2\pi)^3} \frac{M}{k^2-q^2+i\epsilon} 
= {} - \frac{M}{4\pi} (\mu+ik) \ .
\label{loop}
\eeq
This introduces the renormalization scale $\mu$ into amplitudes,
and so the low-energy constants in Eq.~(\ref{Leff}) must also depend
on $\mu$ to ensure observables are $\mu$-independent.

Note that $\mu$ should be of the same order as
the characteristic momentum of the process, $k$, so that the relative
size of the loop Eq.~(\ref{loop}) is not altered.
Both scales should be small compared to the scale of
underlying physics ($m_\pi$ for neutrons)
to ensure the EFT description does not break down.
Furthermore, two-particle interactions
exhibiting a large scattering length introduce a small momentum scale
into the system (the neutron scattering length $\as$ 
gives $|1/\as|\sim10$~MeV).
All this can be summarized by defining a generic small scale
$Q\equiv(k,\mu,1/\as)$, and 
all observables are well
approximated by an expansion in $Q$.

Matching the on-shell T-matrix to phase-shift data
gives expressions for the low-energy constants of
Eq.~(\ref{Leff}) in terms of scattering observables~\cite{Kaplan:1998we}
such as the scattering length $\as$ and effective range $r_e$,
\ben
C_0 = - \frac{4\pi/M}{\mu-1/\as} \sim \frac1Q \ ,
\qquad
C_2 \, k^2 = \frac{4\pi/M}{(\mu-1/\as)^2} \frac{r_e}2\; k^2 \sim Q^0 \ .
\een
Since the particle-particle loop Eq.~(\ref{loop}) scales like $Q$, all
insertions of $C_0$ contribute at leading order $1/Q$ to the
T-matrix,
whereas insertions of $C_2$ and other contact interactions are higher
order.
This justifies the Lippmann-Schwinger equation for $C_0$ only, which
forms a summable geometric series, giving~\cite{Kaplan:1998we}
\beq
T(k) = \frac{C_0}{1-\frac{M C_0}{4\pi} (\mu+ik)} =
\frac{4\pi/M}{1/\as+ik} \ .
\eeq
Note that the result is $\mu$-independent and produces the first term
in the effective range expansion by construction.

A many-body system of fermions with spin-isospin 
degeneracy $g=(2S+1)(2I+1)$ 
forms a Fermi sea with
Fermi momentum $\kf$ and density $\rho=g\kf^3/6\pi^3$.
The kinetic energy per particle
of a free fermion gas 
is given by the familiar expression
\beq
\frac{E_{\rm kin}}{A} = \frac{g}{\rho} \int\! \frac{d^3k}{(2\pi)^3} 
\frac{k^2}{2M} = \frac35 \frac{k_F^2}{2M} \ .
\label{kin}
\eeq
Quantum fluctuations allow particle-hole pairs to form,
with interactions still given by the lagrangian Eq.~(\ref{Leff}).
These contributions to the ground-state energy can be represented by
closed, connected Feynman diagrams and calculated using finite-density
Feynman rules~\cite{Negele,Hammer:2000xg}. 

Two basic building blocks of finite-density diagrams are depicted in
Fig.~\ref{fig1}a and b.
Since the neutron propagator accounts for both particle and hole
propagation 
\beq
G_0(k)_{\alpha\beta} = \delta_{\alpha\beta} 
\( \frac{\theta(k-k_F)}{k_0-{\bf k}^2/2M+i \epsilon}
+ \frac{\theta(k_F-k)}{k_0-{\bf k}^2/2M-i \epsilon}
\) \ ,
\label{g0}
\eeq 
contour integration of the loop momentum shows Fig.~\ref{fig1}a and b
actually represent {\em three} processes:
two-particle or two-hole scattering in Fig.~\ref{fig1}a
and one-particle, one-hole scattering in Fig.~\ref{fig1}b.
Pauli blocking is enforced by the following theta functions,
which depend on the center of mass momentum 
${\bf P}={\bf k}_1+{\bf k}_2$ 
and arbitrary relative momentum ${\bf q}$ (with 
${\bf q}_{{\bf 1},{\bf 2}} =
\frac12 {\bf P}\pm {\bf q}$), 
\ben
\theta_q^+ &\equiv& \theta\!\( q_1-\kf\) \,
\theta\!\( q_2-\kf\) \ ,
\\
\theta_q^\pm &\equiv& \theta\!\( q_1 -\kf\) \,
\theta\!\( \kf-q_2\) \ ,
\\
\theta_q^- &\equiv& \theta\!\( \kf - q_1\) \,
\theta\!\( \kf-q_2\) \ ,
\een
for particle-particle, particle-hole,
and hole-hole scattering respectively.
These three processes can be integrated analytically: 
\be
&&\int\!\! \frac{d^3q}{(2\pi)^3} \;\frac{\theta_q^+}{k^2-q^2+i\epsilon} 
= - \frac{\mu}{4\pi} + \frac{\kf}{(2\pi)^2} f(\kappa,s) \ ,
\label{pp}
\\
&&\int\!\! \frac{d^3q}{(2\pi)^3} \;\frac{\theta_q^\pm}{{\bf P} \cdot ({\bf
k}-{\bf q}) +i\epsilon} 
= -\frac{\kf}{(2\pi)^2} 
\wt f(\kappab \cdot {\bf s},s) 
\ ,
\label{ph}
\\
&&\int\!\! \frac{d^3q}{(2\pi)^3} \;\frac{\theta_q^-}{q^2-k^2+i\epsilon} 
= \frac{\kf}{(2\pi)^2} f(\kappa,-s) \; \theta (1-s) \ .
\label{hh}
\ee

\noindent
having defined the dimensionless momenta ${\bf s} = \frac12 {\bf P}/\kf$ 
and $\kappab = {\bf k}/\kf$.
Particle-particle scattering has contributions from momenta up to
infinity
and so can be written as a regularized integration over all momenta,
Eq.~(\ref{loop}), which brings the $\mu$-dependence in Eq.~(\ref{pp}), 
minus a term constraining at least one
propagator to be inside the Fermi surface,
which replaces the remnant of asymptotic scattering, $ik$,
by the nontrivial function $f(\kappa,s)$.
This same functional form appears in hole-hole scattering, 
but in this case both ${\bf k}_1$ and ${\bf k}_2$ must be inside the
Fermi surface, and so 
$P<2\kf$ (i.e.\ $s<1$).
In fact, the functions in Eqs.~(\ref{pp}-\ref{hh}) have different analytical
forms depending on whether $s>1$ or $s<1$,
and the combinations which enter ground-state energy calculations are
\be
\theta(1-s)\; f(\kappa,s) &=& 1+s+
\kappa \ln \left|
\frac{1+s-\kappa}{1+s+\kappa}\right|
+ \frac{1-\kappa^2-s^2}{2s} \ln \left|
\frac{(1+s)^2-\kappa^2}{1-\kappa^2-s^2}\right| \ ,
\label{f0}
\\
\theta(1-s) \; \wt f(\alpha s,s) &=& \frac{1-\alpha}2 
+ \alpha \ln \left|\frac{1-s-\alpha}{\alpha} \right| 
+ \frac{1-(\alpha-s)^2}{4s} 
\ln \left| \frac{1+s-\alpha}{1-s-\alpha} \right| \ ,
\label{f1}
\\
\theta(s-1) \; \wt f(\alpha s,s) &=& \frac{s-\alpha}{2s} 
+\frac{1-(\alpha-s)^2}{4s} \ln
\left| \frac{s+1-\alpha}{s-1-\alpha} \right|  \ .
\label{f2}
\ee
Interpreting one contribution to the ground-state energy, Fig.~\ref{fig1}d,
in three different ways gives the identities:
\be
&&\int\! \frac{d^3k\, d^3P}{(2\pi)^6} \;\theta_k^- 
\( f(\kappa,s) - \frac{\pi\mu}{\kf} \)
=
-\int\! \frac{d^3k\, d^3P}{(2\pi)^6} \;\theta_k^\mp 
\wt f(\kappab\cdot {\bf s},s)  
\nonumber\\
&&=
\int\! \frac{d^3k\, d^3P}{(2\pi)^6} \;\theta_k^+ 
f(\kappa,-s)  \;\theta(1-s) 
= \frac{\kf^6}{6\pi^4} \( \frac{11-2\ln 2}{35} -
\frac{\pi\mu}{6\kf} \) \ ,
\label{iden}
\ee
which all integrate to the same result, serving
as an excellent check of 
Eqs.~(\ref{f0}-\ref{f2}).

Extending $Q$-counting to finite density shows
the small parameter is $Q\equiv(\kf,\mu,1/\as)$,
so all powers of $\kf\as$ must be summed at leading
order~\cite{Furnstahl:2000ix}.
Also, this choice $\mu\sim\kf$ ensures the regularization does not
destroy the relative importance of Eq.~(\ref{pp}), 
since $f(\kappa,s)/\pi\sim\order(1)$. 
With this scaling, the kinetic energy per particle Eq.~(\ref{kin})
is $\order(Q^2)$, and the contribution with one $C_0$ insertion,
Fig.~\ref{fig1}c, is
\beq
\( \frac{E}{A} \)_{(\ref{fig1}c)} =
\frac{g(g-1)}{2\rho} \int\!\frac{d^3k\, d^3P}{(2\pi)^6} \theta_k^- C_0 
= \frac{(g-1) k_F^3}{12\pi^2} C_0 \sim Q^2 \ .
\label{eoa}
\eeq
The free-space counting $C_0\sim1/Q$ makes Eq.~(\ref{eoa})
of the same order as the kinetic energy and hence 
nontrivial properties such as nuclear binding 
can possibly occur at leading order.
In fact, 
each additional $C_0$ vertex added to Fig.~\ref{fig1}c
requires two more propagators and one more loop integral,
bringing an extra factor of order%
\footnote{The identification $k_0\sim k^2/2M$ is made for power
counting purposes~\cite{Kaplan:1998we}.}
\beq
C_0 \( \frac{M}{\kf^2} \)^2 \( \frac{\kf^5}{M} \) \sim Q^0
\eeq
and so {\em all} possible insertions of $C_0$, including
non-planar graphs, are leading order,
making calculations intractable.


\begin{figure}[t]
\begin{center}
\leavevmode
\epsfysize=1in
\epsffile[255 541 376 628]{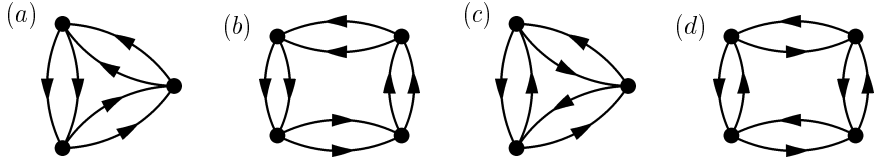}
\end{center}
\caption{\label{fig2} Four representative contributions to the ground-state
energy.
} 
\end{figure}

However, evaluation of some representative diagrams will show there is
still an additional organizational scheme which can be 
exploited.
Take, for example, the Feynman diagrams in Fig.~\ref{fig2}a-d,
which are all of the same order in $Q$-counting.
They are rings of bubbles with particle scattering 
in either the same direction or opposite directions.
A loop momentum is assigned to each bubble and a propagator
Eq.~(\ref{g0}) to each line. 
The final loop
momentum can be chosen to 
be the center-of-mass momentum $P$, which traverses the
entire ring and is shared equally by the legs of each bubble.
A contour integration over the propagators 
enforces energy conservation, so not all propagators of a
diagram can be particles (or holes). 

The contributions to each
diagram are:  
two particle-particle bubbles, one hole-hole bubble (2pp-1hh) and
one particle-particle bubble, two hole-hole bubbles (1pp-2hh) for
Fig.~\ref{fig2}a;
3pp-1hh, 2pp-2hh, and 1pp-3hh for Fig.~\ref{fig2}b;
2ph-1hp and 1ph-2hp for Fig.~\ref{fig2}c; 
3ph-1hp, 2ph-2hp, and 1ph-3hp for Fig.~\ref{fig2}d.
After working through the Feynman rules, 
simplifying the expressions, and
dividing by the density $\rho$, the energy per particle for
each diagram, organized according to the aforementioned particle and
hole content, is:
\ben
&&\( \frac{E}{A} \)_{\!(\ref{fig2}a)}
= (g-1) \frac{\kf^2}{2M} 
\( \frac{M\kf C_0}{4\pi} \)^{\!\!\!3} 
\frac{48}{\pi^3}
\int\! \frac{d^3 s\, d^3\kappa}{(4\pi)^2}  
\[ \theta_\kappa^-
\( f(\kappa,s) - \frac{\pi\mu}{\kf} \)^2 
+ \theta_\kappa^+  f(\kappa,-s)^2 \] \ ,
\\
&&\(\frac{E}{A} \)_{\!(\ref{fig2}b)}
= (g-1) \frac{\kf^2}{2M}
\( \frac{M \kf C_0}{4\pi} \)^{\!\!\!4}
\frac{48}{\pi^4} \int\! \frac{d^3s \, d^3\kappa}{(4\pi)^2}
\\
&&\qquad {}\times 
\[
\theta_\kappa^- \( f(\kappa,s) -\frac{\pi\mu}{\kf} \)^3
+ 3 \, \(f(\kappa,s) -\frac{\pi\mu}{\kf} \)
\int \! \frac{d^3q}{2\pi}
\frac{\theta_\kappa^- \, \theta_q^+}{\kappa^2-q^2} 
f(q,-s)
+ \theta_\kappa^+  f(\kappa,-s)^3
\right] \ ,
\\
&&\( \frac{E}{A} \)_{\!(\ref{fig2}c)}
\!\!\!
= (g-1)(g-3) \frac{\kf^2}{2M} 
\( \frac{M \kf C_0}{4\pi} \)^{\!\!\!3} \frac{96}{2!\, \pi^3} 
\int\! \frac{d^3s\, d^3\kappa}{(4\pi)^2}
\[ \theta_\kappa^{\mp}\, \wt f(\kappab\cdot {\bf s}, s)^2
+\theta_\kappa^{\pm} \, \wt f(-\kappab\cdot {\bf s}, s)^2 \] \ ,
\\
&&\( \frac{E}{A} \)_{\!(\ref{fig2}d)}
= -(g-1)^2 (g-3) \frac{\kf^2}{2M} 
\( \frac{M \kf C_0}{4\pi} \)^{\!\!\!4} \frac{96}{3!\, \pi^4} 
\int\! \frac{d^3s\, d^3\kappa}{(4\pi)^2}
\\
&&\qquad{}\times \[
\theta_\kappa^{\mp} \wt f(\kappab\cdot {\bf s}, s)^3
- 3 \, 
\wt f(\kappab \cdot {\bf s},s)
\int\!\frac{d^3q}{4\pi} \;\frac{\theta_\kappa^\mp\,\theta_q^\pm}
{{\bf s}\cdot (\kappab -
{\bf q} )} \wt f(-{\bf q}\cdot {\bf s},s)
+\theta_\kappa^{\pm} \wt f(-\kappab\cdot {\bf s}, s)^3
\] \ ,
\een
where a factor of $\theta(1-s)$ has been absorbed into $f(\kappa,-s)$.
The factor of ``3'' in contributions to Fig.~\ref{fig2}b and d comes
from the combinatorics of choosing the second pp or ph pair respectively
and the symmetry factor of Fig.~\ref{fig2}d gives an additional $1/3$
suppression compared to the other diagrams.

A simple change of variables shows the $i$ph-$j$hp and
$j$ph-$i$hp contributions are the same for any $i$ and $j$. 
Numerically integrating the above expressions
and taking $\mu=0$ for clarity gives 
\be 
\(\frac{E}{A}\)_{\!(\ref{fig2}a)}
&=& (g-1) \frac{\kf^2}{2M} (\kf \as)^3 
\( 0.0641 + 0.0115 \) \ ,
\label{n1}
\\
\(\frac{E}{A}\)_{\!(\ref{fig2}b)} 
&=& (g-1) \frac{\kf^2}{2M} (\kf \as)^4 
\( 0.0383 +3 \times0.00247 -0.000683 \) \ ,
\label{n2}
\\
\( \frac{E}{A} \)_{\!(\ref{fig2}c)}
&=& (g-1)(g-3) \frac{\kf^2}{2M} \( \kf \as \)^3 
( 0.0287+0.0287) \ ,
\label{n3}
\\
\( \frac{E}{A} \)_{\!(\ref{fig2}d)} 
&=& -(g-1)^2 (g-3) \frac{\kf^2}{2M} \( \kf \as \)^4 
( 0.000603+ 3\times 0.000557+0.000603 ) \ .
\label{n4}
\ee
Adding the terms for each of the three-bubble diagrams,
Fig.~\ref{fig2}a and c, 
gives results in agreement with Ref.~\cite{Hammer:2000xg}.
Comparison of individual terms in Eqs.~(\ref{n1}-\ref{n4}) shows
each additional hole-line is suppressed
by about a factor of three.
This means every added hole-hole bubble 
is suppressed by almost an order of magnitude.
The suppression is the same at nonzero $\mu$, where $\kf\as$ can be
taken very large.


This suppression can be described by a new
power counting at finite density which is kinematical in nature
and will lead to the simplification of many-body calculations.
Take the 1pp-1hh contribution of
Eq.~(\ref{iden}), which using Eq.~(\ref{pp}) can be written in terms
of one relative momentum below the Fermi surface and another above:
\beq
\int\!\frac{d^3P\, d^3k\, d^3q}{(2\pi)^9} 
\frac{\theta_k^-\, \theta_q^+}{k^2 - q^2 + i \epsilon} 
= 
\int\!\frac{d^3P\, d^3k\, d^3q}{(2\pi)^9} 
\frac{\theta_k^-\, \theta_q^+}{- q^2 + i \epsilon} 
\( 1- \frac{k^2}{-q^2+i\epsilon} + \order(k^4/q^4)
\)
\ .
\label{example}
\eeq
The integration region $\theta_k^-$ can be pictured
as the overlap of two Fermi spheres (of radius $\kf$) with centers
separated by the center-of-mass momentum ${\bf P}$,
whereas the region $\theta_q^+$ is given by all momenta excluding
these same intersecting spheres~\cite{Negele}.
In principle, the momenta $k$ and $q$ can both be near
$\kf$, but the integration measure $d^3P$ suppresses $P\sim0$, which makes
the approximation $q\gg k$ appropriate.

In addition, for $P\simlt 2\kf$, the overlap region of the two spheres
is small and suppresses additional powers of $k^2$ in the numerator,
since 
\be
&&\int\!\frac{d^3\kappa}{4\pi} \; \frac{\theta_\kappa^+}{-\kappa^2+i
\epsilon} 
= - \frac{\pi\mu}{2\kf} + 1 + \order(1-s) \ ,
\label{t1}
\\
&&\int\!\frac{d^3\kappa}{4\pi}\; \theta_\kappa^-  =
 \frac{(1-s)^2}{2} +\order(1-s)^3 \ ,
\label{t2}
\ee
and more generally $\theta_q^+/q^{2n}\sim\order(1-s)^0$ for every $n$,
and $\theta_k^-\, k^{2n}\sim\order(1-s)^{n+2}$.
Then, working in $D$ dimensions 
(with the understanding that $D=4$ will
be taken at the end of the calculation) 
the integral over the center-of-mass
momentum gives
\beq
\int_0^1 ds\, s^{D-2} (1-s)^n 
\sim \frac1{D^n} \int_0^1 ds\, s^{D-2}\ .
\label{ood}
\eeq
for large $D$.
In fact, carrying out the entire calculation in $D$-dimensions shows
the suppression is actually 
$1/\cD^n$ with ${\cal D}\equiv2^{D/2}$, which is numerically
equivalent to Eq.~(\ref{ood})
for $D=4$ but even more convergent in larger number of dimensions.

From the above, the theta functions which enforce Pauli blocking
can be associated with powers of $1/\cD$: $\theta_\kappa^+\sim1/\cD^0$,
$\theta_\kappa^-\sim 1/\cD^2$, and 
Eq.~(\ref{iden}) implies $\theta_\kappa^\pm\sim1/\cD$.
Therefore, the expansion in $1/\cD$ is similar to 
the traditional
hole-line expansion of nuclear physics~\cite{Negele}.
Furthermore, the complicated functions of Eq.~(\ref{f0}-\ref{f2}) 
greatly simplify to
\beq
f(\kappa,s) = 2 + \order(\cD^{-1}) \ ,
\qquad
f(\kappa,-s) = - \frac{(1-s)^2}{\kappa^2} + \order(\cD^{-3}) \ ,
\eeq
since they always appear with the respective 
theta functions $\theta_\kappa^-$ and $\theta_\kappa^+$, and
using Eq.~(\ref{iden}),
\beq
\theta(s-1)\; \wt f(\alpha s,s) = \frac1{3s(s-\alpha)} + \order(\cD^{-2}) \ .
\eeq
Using these simplified forms along with Eqs.~(\ref{t1}-\ref{t2}),
the above analytic expressions for the diagrams of Fig.~\ref{fig2}
can be analytically integrated.
The numerical values at each order in $1/\cD$ are collected in
Table~\ref{tab1}, and the $1/\cD$-expansion quickly converges to the
full result in most cases, although the contributions to
Fig.~\ref{fig2}c require further investigation. 

\begin{table}[t] 
\caption{\label{tab1} Exact factors in contributions to the energy
per particle of Eq.~(\ref{n1}-\ref{n4}) and their respective expansion
in powers of $1/\cD$.
An overall factor of ``3'' for the 2pp-2hh and 2ph-2hp terms 
has been left out below for a more direct comparison.  The 1pp-3hh
diagram first appears at order $1/\cD^6$ with contribution $-0.000507$.}
\begin{tabular}{c|cc|ccc|c|cc}
 & \multicolumn{2}{c|}{Fig.~\protect\ref{fig2}a} & 
\multicolumn{3}{c|}{Fig.~\protect\ref{fig2}b} &
Fig.~\protect\ref{fig2}c &
\multicolumn{2}{c}{Fig.~\protect\ref{fig2}d} \\  
   & 2pp-1hh & 1pp-2hh & 3pp-1hh & 2pp-2hh & 
	1pp-3hh & 2ph-1hp & 3ph-1hp & 2ph-2hp  \\  \hline \hline
total & 0.0641\ & 0.0115\ & 
	0.0383\ & 0.00247\  & $-0.000683$ & 
	0.0287\ &
	0.000603\ & 0.000557\  \\ \hline
${\cal O}(1/\cD^2)$ & 0.0860\ & --- & 
	0.0548\ & ---  & --- & 
	--- &
        --- & --- \\
${\cal O}(1/\cD^3)$ & $\!\!\!\!\!-0.0162$ & --- &
	$\!\!\!\!\!-0.0155$ & --- & --- &
	0.0283\ &
	--- & --- \\ 
${\cal O}(1/\cD^4)$ & $\!\!\!\!\!-0.0056$ & 0.0133\ &
	$\!\!\!\!\!-0.0032$  & 0.00310  &  --- &
	0.0057\ &
        0.000676\  & 0.000676\  \\
${\cal O}(1/\cD^5)$ & 0.0002 & $\!\!\!\!\!-0.0024$ & 
	$\!\!\!\!\!-0.0007$  & $\!\!\!\!\!-0.00164$  & --- & 
	$\!\!\!\!\!-0.0003$ &
        0.000371\  & $\!\!\!\!\!-0.000230$  \\
\end{tabular}
\end{table}


The existence of a power counting in $1/\cD$ allows for
systematic calculation of finite-density observables such as the
ground-state energy.
First of all, it is interesting to note that at next-to-leading order in the
$Q$-expansion, 
one insertion of $C_2 k^2$ will appear, 
and the integration of $k^2$ over phase space ($\theta_k^-$)
brings a factor of $1/\cD$.
Therefore the hierarchy of the 
$Q$-expansion is already included in the $1/\cD$-expansion.
The leading order ground-state energy per particle 
can be written in closed form: 
\beq
\frac{E}{A} = \frac{E_{\rm kin}}{A} + \frac{E_{\rm int}^{(0)}}{A} +
\order(1/\cD)  \ ,
\eeq
with the kinetic energy given by Eq.~(\ref{kin}) and
the interaction energy to this order
given by all the particle-particle scattering
terms (i.e., the $n$pp-1hh bubbles for all $n$).
This is a summable geometric series, giving
\beq
\frac{E_{\rm int}^{(0)}}{A} = \frac{g(g-1)}{2\rho}
\int\! \frac{d^3P\, d^3k}{(2\pi)^6} \; \theta_k^- \;
\frac{4\pi/M}{1/\as-\frac{\kf}{\pi} f(\kappa,s)} \ ,
\eeq
which is $\mu$-independent, as it should be for any consistent
expansion. 
Actually, to the order quoted in $1/\cD$, the substitution
$f(\kappa,s)\to2$ is valid, which gives the leading-order 
ground-state energy per particle for neutron matter (taking $g=2$)
\beq
\frac{E}{A} = \(\frac35 + 
\frac{2\kf \as/3}{\pi-2\kf\as} \) \frac{\kf^2}{2M} 
+ \order(1/\cD)
\ .
\label{result}
\eeq
For $\as>0$, $E/A$ has a pole at $\kf\as=\pi/2$ 
and it would be interesting to explore its consequence in atomic
systems where the scattering length can be tuned.
As for neutrons, the scattering length is negative
and it is a good approximation to take $1/\as\to0$,
to arrive at the scale-invariant answer
$E/A=(4/15) \kf^2/2M$.
The next order in $1/\cD$ will require a single insertion of $C_2$ or
$C_2^p$ and diagrams with three hole-lines. 

A new power counting at finite density has been found which includes
the nonperturbative free-space power counting of two-nucleon EFTs. 
The many-body problem simplifies 
in a large number of space-time dimensions and 
leads to a convergent expansion even for $D=4$.
Including explicit pions and three-body forces in the lagrangian will
allow the above analysis to be extended to nuclear matter 
for an exploration of the behavior near saturation density.


I would like to thank R.~J.~Furnstahl, H.~W.~Hammer,
D.~B.~Kaplan, J.~W.~Negele, and K.~Rajagopal for useful discussions
and I.~W.~Stewart for a critical reading of the manuscript. 
This work was supported in part by the U.S. Department of Energy under
cooperative research agreement \#DF-FC02-94ER40818.

\end{document}